\documentclass[amsthm]{autart}

\usepackage[cmex10]{amsmath}
\usepackage{amssymb}
\usepackage{txfonts}
\usepackage{array}
\usepackage{accents}
\makeatletter
  \def\widebar{\accentset{{\cc@style\underline{\mskip10mu}}}}
  \def\wideubar{\underaccent{{\cc@style\underline{\mskip10mu}}}}
\makeatother
\usepackage[dvipdfmx]{graphicx} %

\usepackage{cite}
\usepackage[labelformat=simple]{subcaption}

\makeatletter
\renewenvironment{pf}%
  {\par%
   {\bfseries\Elproofname}\enspace\ignorespaces}%
  {$\quad\square$\par}%
\def\Elproofname{Proof.}
\@namedef{pf*}#1{\par\begingroup\def\Elproofname{#1}\pf\endgroup\ignorespaces}
\expandafter\let\csname endpf*\endcsname=\endpf
\makeatother

\newcommand{\R}{\mathbb{R}}
\newcommand{\Zp}{\mathbb{Z}_+}
\newcommand{\N}{\mathbb{N}}
\newcommand{\E}{\textnormal{E}}
\newcommand{\Prob}{\textnormal{Prob}}
\newcommand{\diag}{\textnormal{diag}}

\newcommand{\calY}{\mathcal{Y}}
\newcommand{\bY}{\widebar \calY}
\newcommand{\uY}{\wideubar \calY}

\newcommand{\hcalY}{\hat \calY}
\newcommand{\calA}{\mathcal{A}}
\newcommand{\calB}{\mathcal{B}}

\newcommand{\Rnec}{R_{\textnormal{nec}}}%
\newcommand{\Rsuf}{R_{\textnormal{suf}}}
\newcommand{\pl}{p}
\newcommand{\pr}{q}

\newcommand{\prnec}{q_{\textnormal{nec}}}%
\newcommand{\Nnece}{N^{(\textnormal{e})}}
\newcommand{\Nneco}{N^{(\textnormal{o})}}
\newcommand{\bw}{\widebar w}
\newcommand{\bwzero}{\widebar w^{(0)}}
\newcommand{\bwone}{\widebar w^{(1)}}
\newcommand{\ud}[1]{\Delta_{u_{#1}}}%

\usepackage{mathtools}
\DeclarePairedDelimiter\ceil{\lceil}{\rceil}

\newcommand{\fref}[1]{Fig.~\ref{#1}}

\newcommand{\IEEEJAC}{{IEEE} Trans. Autom. Control}
\newcommand{\SIAMCO}{SIAM J. Control Optim.}
\newcommand{\SysCL}{Syst. Control Lett.}
\newcommand{\IEEEJPROC}{Proc. {IEEE}}
\newcommand{\IJRN}{Int. J. Robust Nonlin. Control}
\newcommand{\IJACSP}{Int. J. Adapt. Control Signal Process.}

\begin{document}
\begin{frontmatter}
\title{Stabilization of uncertain systems using quantized and\\ 
lossy observations and uncertain control inputs\thanksref{footnoteinfo}}
\thanks[footnoteinfo]{%
This work was supported in part by the JST-CREST Program,
by JSPS Postdoctoral Fellowship for Research Abroad,
and by JSPS KAKENHI Grant Number JP16H07234.}

\author[Okano]{Kunihisa Okano}\ead{kokano@okayama-u.ac.jp},
\author[Ishii]{Hideaki Ishii\corauthref{cor}}\ead{ishii@c.titech.ac.jp}
\corauth[cor]{Corresponding author.}
\address[Okano]{Department of Intelligent Mechanical Systems,
  Okayama University, Okayama, 700-8530, Japan}
\address[Ishii]{Department of Computer Science,
Tokyo Institute of Technology,
Yokohama, 226-8502, Japan}

\begin{abstract}
In this paper, we consider a stabilization problem of an uncertain system
in a networked control setting. 
Due to the network, the measurements are quantized to finite-bit signals and may
be randomly lost in the communication.
We study uncertain autoregressive systems whose state and input parameters vary
within given intervals.
We derive conditions for making the plant output to be mean square stable,
characterizing limitations on data rate, packet loss probabilities, and
magnitudes of uncertainty.
It is shown that a specific class of nonuniform quantizers can achieve
stability with a lower data rate compared with the common uniform one.
\end{abstract}
\end{frontmatter}

\section{Introduction}
This paper studies stabilization of a linear system in which the plant outputs
are transmitted to the controller through a bandwidth limited lossy channel and
the exact plant model is unavailable.
For control over finite data rate channels, it is well known
\cite{Tatikonda2004, Nair2004} that there exists a tight bound on the data rate
for stabilization of linear systems, which is expressed simply by the product of
the unstable poles of the plant.
Such data rate limitations have been developed under a variety of  networked
control problems.
For general nonlinear systems, it has been pointed out that the limitation is
related to topological entropy \cite{Nair2004a, Savkin2006, Colonius2013}.
For an overview on the topic, we refer to \cite{Nair2007};
for more recent works, see, e.g., \cite{Heemels2010, Yuksel2014}.
On the other hand, control over packet dropping channels has also been 
studied actively (see, e.g., \cite{Hespanha2007,Schenato2007}).
Interestingly, by modeling the behavior of the losses as  i.i.d.\ random
processes, the maximum packet loss probability for achieving stabilization can
also be characterized solely by the product of the unstable poles of the plant. 
Recent works have extended such results to the case of Markovian packet losses.
In particular, \cite{You2011a} and \cite{Minero2013} have derived the minimum
data rates for the static and time-varying rate cases, respectively.
We note that the works mentioned above assume perfect knowledge of the plant
models.

Despite the active research in the area of networked control, uncertainties in
plant models have received limited attention and are thus the focus of this work.
In general, it is difficult to deal with the combination of uncertainties in the
systems and incompleteness in the communication.
In particular, in data rate limited control problems, the state evolution must
be estimated through quantized information, but in the uncertain case, this task
becomes complicated and often conservative. 
In \cite{Phat2004, Fu2010}, linear time-invariant systems with norm bounded
uncertainties are considered, and controllers to robustly stabilize the systems
are proposed.
In \cite{Martins2006}, scalar nonlinear systems with stochastic uncertainties
and disturbances are studied, and a data rate bound sufficient for the moment
stability is derived.
Moreover, related stabilization problems are studied from the viewpoints of 
adaptive control \cite{Hayakawa2009} and switching control \cite{Vu2012}
as well.
In these results, however, only sufficient conditions on data rates have been
obtained, and they are not concerned with characterizing the minimum.
On the other hand, observation problems of nonlinear time-varying uncertain
systems are studied in \cite{Savkin2006}.
Both required and sufficient data rates for observability are characterized by
using the notion of topological entropy.

More specifically, we consider the stabilization of a parametrically uncertain
plant over a Markovian lossy channel. 
The plant is represented as an autoregressive system whose parameters vary
within given intervals.
We develop bounds on the data rate, the packet loss probability, and plant
uncertainty for stabilizability.
The results become tight for the scalar plants case.
In the course of our analysis, we demonstrate that the data rate
can be minimized by employing a class of nonuniform quantizers, which is
constructed in an explicit form.
These quantizers have an interesting property that the cells are coarser around
the origin and finer further away from the origin; such a structure is in
contrast to the well-known logarithmic quantizer \cite{Elia2001}.

For the case of uncertain state parameters, the authors have studied the minimum
data rate under model uncertainties and packet losses for the i.i.d.\
\cite{Okano2012b} and the Markovian \cite{Okano2014a} lossy channels. 
It is also noted that, in \cite{Kang2015}, for a similar class of uncertain
plants, stabilization techniques have been developed based on the logarithmic
quantizers \cite{Elia2001}.
This paper aims at further studying the more realistic situation where
uncertainty is also present in the actuator of the plant.
That is, the parameter of the control input may also be uncertain.

The main difficulty in the current setup can be described as follows.
Evaluation of the estimation error and its evolution is the key to derive
the minimum data rate.
Due to plant instability, the estimation error grows over time, but it can be
reduced based on state observations.
In our previous work \cite{Okano2014a}, we have assumed uncertainty only in the
state coefficients.
In the presence of uncertainty in the actuator, the results there are not
applicable.
In particular, when the control input is large, the estimation error will grow
further, making the analysis more involved.
We show that uncertainty in the actuator side introduces additional
nonuniformity in the quantizer structure when compared to our previous results. 

This paper is organized as follows.
In the next section, we formulate the stabilization problem for the networked
control system.
In Section~\ref{sec,scalar}, we consider the fundamental case for the scalar
plant systems.
The general order plants case is considered and a sufficient condition for the
stability is shown in Section~\ref{sec,multi}.
In \cite{Okano2014a}, a necessary condition for the general order plants case is also provided.
However, in this paper, we do not include the corresponding result since
in general it contains some conservativeness and its significance may be limited.
Finally, concluding remarks are given in Section~\ref{sec,conclusion}.
The material of this paper was presented in \cite{Okano2014b} in a preliminary
form, but this version contains updated results with their full proofs.

{\em Notations:}
$\Zp$ is the set of nonnegative integers and $\N$ stands for the set of natural
numbers.
$\log_2(\cdot)$ is simply written as $\log(\cdot)$.
For a given interval $\calY$ on $\R$, 
denote its infimum, supremum, and midpoint by $\uY$, $\bY$, and
$c(\calY):=(\uY+\bY)/2$, respectively;
its width is given by $\mu(\calY):=\bY-\uY$.

\section{Problem setup}\label{sec,problem}
We consider the networked system depicted in \fref{fig,system}, where
the plant is connected with the controller by the communication channel.
At time $k\in\Zp$, %
the encoder observes the plant output $y_k\in\R$ and quantizes it to a
discrete value.
The quantized signal $s_k\in\Sigma_N$ is transmitted to the decoder through
the channel.
Here, the set $\Sigma_N$ represents all possible outputs of the encoder
and contains $N$ symbols.
Thus, the data rate $R$ of the channel is given as $R:=\log N$;
thus, the term refers to the bit rate of the communication, which is
consistent with the literature \cite{You2011a, Minero2013}.
The decoder receives the symbol and decodes it into the interval
$\calY_k\subset \R$, which is an estimate of $y_k$.
The transmitted signal $s_k$ may be lost in the channel.
The result of the communication is notified to the encoder by the
acknowledgment signals before the next communication starts.
Finally, using the past and current estimates, the controller provides the
control input $u_k\in\R$.
\begin{figure}[t]
\centering
\includegraphics[scale=1]{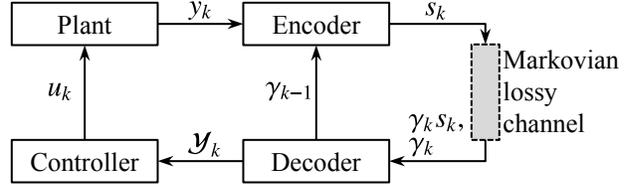}
\caption{Networked control system}
\label{fig,system}
\end{figure}

The plant is the following autoregressive system with uncertain parameters:
\begin{align}
 y_{k+1}\!=\!a_{1,k}y_k\!+\!a_{2,k}y_{k-1}\!+\!\cdots\!+\!a_{n,k}y_{k-n+1}+b_ku_k.\label{AR}
\end{align}
The parameters are bounded and may be time varying as
\begin{align}
 &a_{i,k}\in\calA_i:=\left[a_i^*-\epsilon_i, a_i^*+\epsilon_i\right],\quad
  i=1, 2, \dots, n,\notag\\
 &b_k\in\calB:=\left[b^*-\delta,b^*+\delta\right],\label{uncertainty}
\end{align}
where $\epsilon_i\geq0$ and $\delta\geq0$.
The initial values $y_k$, $k=-n+1, \dots, -1, 0$, are in the known intervals
as $y_k\in Y_k$, where $0<\mu(Y_k)<\infty$.
To ensure controllability at all times, we introduce the following assumption:
For every time $k\in\Zp$, the input parameter $b_k$ is nonzero.
That is, 
\begin{align}
  |b^*|-\delta>0.\label{b-d>0}
\end{align}
In the communication channel, transmitted signals may be randomly lost.
Denote the channel state at time $k$ by the Markovian random variable
$\gamma_k\in\{0,1\}$.
This state represents whether the packet is received ($\gamma_k=1$)
or lost ($\gamma_k=0$).
The loss probability at time $k$ depends on the previous state $\gamma_{k-1}$
and is denoted by the failure probability $\pl$ and the recovery probability
$\pr$ as follows:
$\Prob(\gamma_k\!=\!0\, |\, \gamma_{k-1}\!=\!0)=1-\pr,\
 \Prob(\gamma_k\!=\!1\, |\, \gamma_{k-1}\!=\!0)=\pr,\
 \Prob(\gamma_k\!=\!0\, |\, \gamma_{k-1}\!=\!1)=\pl,\
 \Prob(\gamma_k\!=\!1\, |\, \gamma_{k-1}\!=\!1)=1-\pl$.
To make the process $\{\gamma_k\}_k$ ergodic, assume $\pl,\pr\in(0,1)$.
Moreover, without loss of generality,
assume that at the initial time $k=0$ the packet is successfully transmitted,
i.e., $\gamma_0=1$.
The communicated signal $s_k\in\Sigma_N:=\{1,2,\dots,N\}$ is the quantized
value of the plant output $y_k$ and
is generated by the encoder as $s_k=\phi_N(y_k/\sigma_k)$.
Here, the quantizer $\phi_N(\cdot)$ is a time-invariant map from
$[-1/2,1/2]$ to $\Sigma_N$ with a scaling parameter $\sigma_k>0$.
The quantizer divides its input range $[-1/2,1/2]$ into $N$ cells and its
output is the index of the cell into which the input falls.
We assume the boundaries of the quantization cells to be symmetric about
the origin.
When $N$ is even, we denote the boundary points of nonnegative
quantization cells as $h_{l}$, $l=0,1,\dots,\lceil N/2\rceil$, where
\begin{align}
 h_0=0,\quad  h_{\lceil N/2\rceil}=\frac{1}{2},\quad h_l<h_{l+1}.
 \label{quantizeredge}
\end{align}
If $N$ is odd, the nonnegative boundaries can be written by $h_l$,
$l=1,2,\dots,\ceil{N/2}$, where $0<h_1<h_2<\cdots<h_{\ceil{N/2}}=1/2$.
For notational simplicity, we add $h_0=0$ and use the same notation
$\{h_l\}_{l=0}^{\ceil{N/2}}$ for this case also.

Based on the channel output $\gamma_k s_k$, the decoder determines the
interval $\calY_k\subset\R$, which is the estimation set of $y_k$.
When the packet arrives successfully, i.e., $\gamma_k=1$,
$\calY_k$ is the quantization cell which $y_k$ falls in.
Otherwise, $\calY_k$ is taken as the entire input range $[-\sigma_k/2,\sigma_k/2]$
of the quantizer.

The initial value of the scaling parameter $\sigma_k$ and its update law
are shared between the encoder and the decoder.
The parameter $\sigma_k$ is updated as follows.
At time $k$, the encoder and the decoder predict the next plant output
$y_{k+1}$ based on the past estimates $\calY_{k-i+1}$, $i=1,2,\dots,n$.
As time progresses from $k$ to $k+1$, the past outputs
$y_{k-i+1}\in\calY_{k-i+1}$ are multiplied by $a_{i,k}\in\calA_i$.
Let $\calY_{k+1}^-\subset\R$ be the set of predicted values for $y_{k+1}$
in the form of $\sum_{i=1}^na_{i,k}y_{k-i+1}$ and is given by
\begin{align}
 \calY^-_{k+1} := \{&a_1y'_k + \cdots + a_ny'_{k-n+1}:
 a_1\in\calA_1,\dots, a_n\in\calA_n,\notag\\
 &y'_k\in\calY_{k}, \dots, y'_{k-n+1}\in\calY_{k-n+1}
 \}.\label{def,calY^-}
\end{align}
Moreover, 
since the applied input is $b_ku_k$, the set 
$\{y^-+bu_k:y^-\in\calY^-_{k+1},b\in\calB\}$
is large enough to include $y_{k+1}$.
Note that the above prediction set in (\ref{def,calY^-}) is computable on
both sides of the channel by the acknowledgment signal
regarding $\gamma_{k-1}$ from the decoder to the encoder.
Finally, this set $\{y^-+bu_k:y^-\in\calY^-_{k+1},b\in\calB\}$ must be covered
by the quantizer to avoid saturation.
Hence, the scaling parameter $\sigma_k$ is a function of $\calY_{k+1}^-$ and
must be large enough that
\begin{align}
 \sigma_{k+1}\geq2\sup_{y^-\in\calY^-_{k+1},\ b\in\calB}|y^-+bu_k|.
 \label{sigma_ineqb}
\end{align}
The controller provides the control input $u_k$ based on the past
and current estimates $\calY_{k-n+1},\dots,\calY_{k}$ as
 \begin{align}
 u_k=\sum^n_{i=1}f_{i,k}\left(\calY_{k-i+1}\right),\label{controller}
 \end{align}
where $f_{i,k}(\cdot)$ are maps from an interval in $\R$ to a real number,
which determine the input based on the estimates.

This paper investigates stabilization of the uncertain networked system in
\fref{fig,system} by designing the encoder, the decoder, and the controller
under the constraints \eqref{quantizeredge}--\eqref{controller}.

\begin{defn}\label{def,stability}
The feedback system depicted in \fref{fig,system} is \emph{stabilizable}
if there exists a pair of an encoder $\phi_N$ with the scaling parameter
$\sigma_k$ satisfying (\ref{sigma_ineqb}) and a controller (\ref{controller})
such that the worst case output $y_k$ over all deterministic perturbations
is mean square stable (MSS):
$\E[\sup_{y\in \calY_k}|y|^2 ]\to 0$ as $k\to\infty$.
Here, $\calY_k$ is the decoder output at time $k$ and the expectation is taken
with respect to the packet losses $\gamma_0$, $\dots$, $\gamma_k$.
\end{defn}

\begin{rem}\label{rem,tcns}
We note that the results and the proofs in this paper have been polished
compared with our previous paper \cite{Okano2014a} thanks to the help of the
anonymous reviewers.
In \cite{Okano2014a}, the definition of stabilizability and the proof
of the necessary condition for the scalar plants case should be updated.
This however does not change the bounds on the data rate and the loss
probabilities shown in the theorem.
\end{rem}

\begin{rem}\label{rem,structures}
To simplify the analysis in dealing with uncertainties, we have introduced
some structures in the encoder and the controller.
While they are closely related to those employed in,
e.g., \cite{Phat2004, Liberzon2005, Martins2006} for obtaining sufficient
conditions, there is some conservatism.
In Definition~\ref{def,stability}, the supremum is taken over $\calY_k$,
which contains all possible $y_k$ over
$\{a_{i,j}\}_{j=0}^k$, $a_{i,j}\in\calA_i$, $i=1,2,\dots,n$,
$\{b_j\}_{j=0}^k$, $b_j\in\calB$, and $y_{j}\in Y_j$, $j=-n+1,-n+2,\dots,0$.
If we do not limit the controller class to \eqref{controller}, there may exist
one which can compute an estimation set tighter than $\calY_k$, though
it is difficult to describe the tightest estimation set analytically.
Furthermore, for Definition~\ref{def,stability}, it is important that
the quantizer does not saturate.
This is guaranteed by \eqref{sigma_ineqb}, and there is always a quantization
cell containing $y_k$.
\end{rem}

\section{Scalar plants case}\label{sec,scalar}
We first analyze the simple setup with the scalar plant (a first-order
autoregressive process):
\begin{align}
 &y_{k+1}=a_ky_k+b_ku_k,\notag\\
 &a_k\in\calA:=\left[a^*\!-\!\epsilon,a^*\!+\!\epsilon\right],\quad
  b_k\in\calB=\left[b^*\!-\!\delta,b^*\!+\!\delta\right],
 \label{scalarplant}
\end{align}
where $\epsilon\geq 0$ and $\delta\geq 0$.
We assume that its dynamics is always unstable in the sense that the parameter
$a_k$ has magnitude greater than 1 at all times, i.e.,
\begin{align}
 |a^*|-\epsilon>1.\label{a-e>1}
\end{align}
To express the main result of this section, let
\begin{align}
 &r_a:=\frac{|a^*|-\epsilon}{|a^*|+\epsilon},\quad
 r_b:=\frac{|b^*|-\delta}{|b^*|+\delta},\quad
 \Delta:=\epsilon+\delta\frac{|a^*|}{|b^*|},\notag\\
 &\nu:=\sqrt{1+\frac{\pl \left\{(|a^*|+\epsilon)^2 - 1\right\}}
            {1-(1-\pr)(|a^*|+\epsilon)^2}}.\label{def,r,Delta,nu}
\end{align}
Here, $r_a$, $r_b$, and $\Delta$ reflect the magnitudes of the uncertainties
and $\nu$ represents the effect of packet losses in the required data rate
as we will see in (\ref{nec,R1}).
We show later, in the proof of the next theorem, that the radicand of $\nu$
is positive when the feedback system is stabilizable.
If $a_k$ is a constant, i.e., $a_k=a$ for all time $k$, it has been shown
in \cite{You2011a} that mean-square stability implies that $q>1-1/a^{2}$.
Taking account of the case that $a=|a^*|+\epsilon$, we have $\nu^2>0$.
The following theorem shows a condition on the data rate $R=\log N$,
the loss probabilities $\pl, \pr$, and the magnitude of uncertainty $\Delta$
for stabilizability.

\begin{thm}\label{th,scalar}
Consider the feedback system in \fref{fig,system} with the scalar plant in
(\ref{scalarplant}).
If the system is stabilizable, then the following inequalities hold:
\begin{align}
 &R > \Rnec
 :=\begin{cases}
    \log\frac{\log\{(1-\Delta\nu)^2\}}{\log (r_a r_b)}
    &\text{if }\epsilon\!>\!0\text{ or }\delta\!>\!0,\\
    \log|a^*|+\log\nu &\text{if }\epsilon=\delta=0,
   \end{cases}\label{nec,R1}\\
 &\pr > \prnec\!:=\!1\!-\!\frac{1}{(|a^*|+\epsilon)^2}
  \!+\!\Delta^2\frac{\pl^2\left\{1-(|a^*|+\epsilon)^{-2}\right\}}{1-\Delta^2},
  \label{nec,p1}\\
 &0 \leq \Delta < 1\label{nec,e1}.
\end{align}
Furthermore, if these inequalities are satisfied with an even $N$, it is
possible to construct a stabilizing controller.
\end{thm}

We note that when \eqref{nec,p1} and \eqref{nec,e1} hold and $\epsilon>0$ or
$\delta>0$, we have that $0<1-\Delta\nu<1$ and hence $\Rnec$ is well defined;
see the last part of the proof of Theorem \ref{th,scalar} for details.
This theorem provides limitations for stabilization on the data rate, the packet
loss probabilities, and the plant uncertainty.
The required data rate $\Rnec$ and the recovery probability $\prnec$ are
monotonically increasing with respect to the uncertainty bounds $\epsilon$
and $\delta$.
This means that more plant uncertainty requires better communication with higher
data rate and recovery probability.

We see that the sum of the uncertainties $\Delta=\epsilon+\delta|a^*|/|b^*|$
appears in the limitations.
It is interesting that there is no explicit limitation on $\epsilon$ or
$\delta$, but the sum $\Delta$ of these uncertainties must be smaller than 1.
This indicates some tradeoff in the tolerable uncertainties for $a_k$ and $b_k$.
In particular, %
the product $\delta|a^*|$ implies that for more unstable plants, the bound
$\delta$ on the input parameter $b_k$ has more effect on the stability
conditions in the theorem. 
We remark that when the input parameter is known and is constant as
$b_k\equiv b^*$, i.e., $\delta=0$, then the limitations $\Rnec$ and $\prnec$
coincide with those shown in \cite{Okano2014a}, where uncertainty is present
only in $a_k$.
Moreover, if $a_k$ is also known, i.e., $\epsilon=\delta=0$, then the
limitations are equal to those in \cite{You2011a}, where the exact plant
model is assumed to be available.

We provide an example to illustrate the limitations in Theorem \ref{th,scalar}.
Consider a plant with $a^*=2.0$ and $b^*=1.0$
and a channel with the loss probabilities $\pl=0.05$ and $\pr=0.90$.
\fref{fig,necR_e_delta} shows the bound $\Rnec$ on the data rate
versus the uncertainties $\epsilon$ in $a_{k}$ and $\delta$ in $b_k$.
When the sum of the uncertainties $\Delta$ is large as (\ref{nec,p1}) is
not satisfied, $1-\Delta\nu$ in \eqref{nec,R1} is nonpositive and hence the
required data rate for stabilizability becomes infinite.
\begin{figure}[t]
 \begin{center}
  \includegraphics[scale=.5]{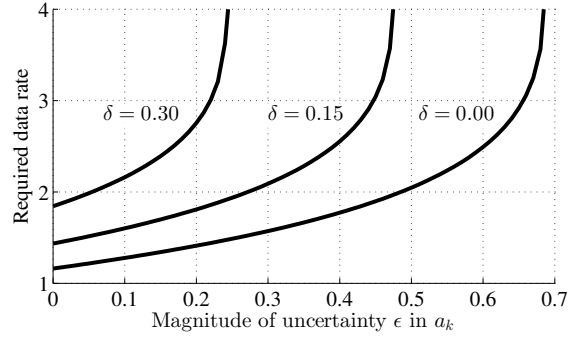}
  \caption{The data rate limitation $\Rnec$ versus the magnitudes of the
   uncertainties $\epsilon$ and $\delta$
   ($a^*=2.0$, $b^*=1.0$, $\pl=0.05$, and $\pr=0.90$)}
  \label{fig,necR_e_delta}
 \end{center}
\end{figure}

\begin{rem}
The works of \cite{Phat2004} and \cite{Martins2006} have shown sufficient
conditions for stabilization of uncertain plants via finite data rate
and \emph{lossless} channels.
We remark that those conditions contain conservatism even for the scalar plants
case.
Consider the scalar plant (\ref{scalarplant}) when the input coefficient is
known ($\delta=0$) and the channel is lossless.
For such systems, the sufficient bound on the data rate $\Rsuf$ in
\cite{Phat2004} and the one $\Rsuf'$ from \cite{Martins2006}, respectively,
become
\begin{align*}
 \Rsuf:=\log\frac{|a^*|-\epsilon(|a^*|+\epsilon)}
 {1-\epsilon(2|a^*|+2\epsilon+1)}~~\text{and}~~
 \Rsuf':=\log\frac{|a^*|}{1-\epsilon}.
\end{align*}
It can be verified that our result is tighter than these bounds as
$\Rnec<R_{\text{suf}}$ and $\Rnec<R_{\text{suf}}'$.
For general order plants, however, it is difficult to compare these results
since the types of uncertainties are different:
In \cite{Phat2004}, unstructured uncertainties are considered, and it is
hard to describe the data rate limitation in an explicit form,
while \cite{Martins2006} deals with nonlinear plants but only scalar ones.
\end{rem}

We now present the proof of Theorem~\ref{th,scalar}.
The key idea %
lies in evaluating the expansion rate of the state estimation sets due to plant
instability.
The proof consists of two steps, which are presented in Sections~\ref{sec,opt_q}
and \ref{proofTh1}, respectively.

\begin{rem}
Compared with our previous work \cite{Okano2014a}, the main difficulty is
that we have to take account of the expansion in the state estimation sets
by control inputs.
If we know the exact control input applied to the plant, then the width of
the estimation set is not affected by the input since we can track the variation
of the state precisely.
However, in the current setup, the estimation set may expand by the control
input due to the uncertainty in $b_k$.
Hence, the scaling parameter $\sigma_{k+1}$ must be selected to cover this
expansion in addition to that by plant instability.
\end{rem}

\subsection{The quantizer minimizing the expansion rate}\label{sec,opt_q}
In this subsection, we introduce the expansion rate for a given quantizer.
Then we show the optimal quantizer which minimizes the rate in the worst case.

For a given quantizer whose boundary points are
$\{h_l\}_{l=0}^{\lceil N/2\rceil}$, let
\begin{align}
 w_{l} \!:=\!
 \begin{cases}
  2(|a^*| \!+\! \epsilon)h_{l+1}
   \hspace{3.5em}\text{if }N\text{ is odd and }l=0,\\
  (|a^*| \!+\! \epsilon)\left(1 \!+\! \frac{\delta}{|b^*|}\right)h_{l+1}
  -(|a^*| \!-\! \epsilon)\left(1 \!-\! \frac{\delta}{|b^*|}\right)h_l\\
  \hspace{8.7em}\text{else},
 \end{cases}\label{def,w}
\end{align}
for $l=0,1,\dots,\lceil N/2\rceil-1$.
We see later in Section \ref{proofTh1} that this $w_l$ characterizes
the expansion rate of the volume of the estimation set for one sampling period
due to the uncertain parameters $a_k\in\calA$ and $b_k\in\calB$.
The rate varies depending on $l$, which represents the cell which observed
output falls into and we have to consider the worst case to guarantee stability
against uncertainties.

The tight lower bound on the worst-case expansion rate can be derived as shown
in Lemma~\ref{lem,LBofw} below.
Let us define the boundary points $\{h_l^*\}_{l=0}^m$ dividing $[-h_m,h_m]$,
where $m\in \left\{ 1,2,\dots, \lceil N/2\rceil\right\}$, as
 \begin{align}
  h^*_{l}:=
  \begin{cases}
   h_m\frac{1-t(r_ar_b)^l}{1-t(r_ar_b)^{m}}
   	&\textrm{if }\epsilon>0 \textrm{ or }\delta>0,\\
   h_m\frac{l-t'}{m-t'}
    &\textrm{if }\epsilon=\delta=0.
  \end{cases}\label{def,h*}
 \end{align}
Furthermore, define $w^*_m$, which is used in the following lemma to represent
the worst-case expansion rate as
\begin{align}
  w^*_m&:=
  \begin{cases}
   h_m \left(|a^*|\!+\!\epsilon\right)
   \left(1\!+\!\frac{\delta}{|b^*|}\right) \frac{1-r_ar_b}{1-t(r_ar_b)^m}
    & \textrm{if } \epsilon>0 \textrm{ or }\delta>0,\\
   h_m\frac{|a^*|}{m-t'} & \textrm{if } \epsilon=\delta=0,
  \end{cases}\notag\\
  t&:= \begin{cases}
  	\frac{1+\delta/|b^*|}{1-\epsilon/|a^*|} &\textrm{if } N \textrm{ is odd},\\
  	1   	&\textrm{if } N \textrm{ is even},
  \end{cases}\notag\\
  t'&:= \begin{cases}
  	\frac{1}{2} &\textrm{if } N \textrm{ is odd},\\
  	0   	&\textrm{if } N \textrm{ is even}.
  \end{cases}\label{def,w*}
\end{align}

\begin{lem}\label{lem,LBofw}
 Given a quantizer $\{h_l\}_{l=0}^{\lceil N/2\rceil}$ dividing $[-1/2,1/2]$
 into $N$ cells, consider a subset of the quantization region $[-h_m,h_m]$,
 where $m\in \left\{ 1,2,\dots, \lceil N/2\rceil\right\}$.
 The worst-case expansion rate of the cells in $[-h_m,h_m]$ is bounded as
 \begin{align}
  \max_{l\in\left\{ 0,1,\dots, m-1\right\}}w_l
  \geq w^*_m.\label{w>=w*}
 \end{align}
 The equality in \eqref{w>=w*} holds if $h_l=h^*_l$ for $l=0,1,\dots,m$.
 Furthermore, consider the subset of the quantization region $[h_{m'},h_m]$,
 where $1\leq m'<m\leq\ceil{N/2}$.
 Then, it follows that
 \begin{align}
  &\max_{l\in\left\{m',\, m'+1,\, \dots,\, m-1\right\}} w_l\notag\\
  &\geq
  \begin{cases}
   \left\{h_m\!-\!(r_ar_b)^{m-m'}h_{m'} \right\}
   \left(|a^*|\!+\!\epsilon\right)\left(1\!+\!\frac{\delta}{|b^*|}\right)
   \!\frac{1-r_ar_b}{1-(r_ar_b)^{m-m'}}\\
    \hspace{11em} \text{if } \epsilon>0 \text{ or }\delta>0,\\
   (h_m-h_{m'})\frac{|a^*|}{m-m'}
    \hspace{5em} \text{if }\epsilon=\delta=0.
  \end{cases}\label{w>=w*m}
 \end{align}
\end{lem}

\begin{pf}
To prove the inequalities \eqref{w>=w*} and \eqref{w>=w*m}, we first assume that
there exists a set of boundary points $\{\hat h_l\}_{l=i}^m$, $i\in\{0,m'\}$, %
such that $\hat h_i=h_i$, $\hat h_m=h_m$, and $w_l$ are the
same for all $l\in\{i,i+1,\dots,m-1\}$, i.e., for a constant $\widehat w$
\begin{align}
 w_l=\widehat w,\quad \forall l\in\{i,i+1,\dots,m-1\}.\label{w_const}
\end{align}
We shall show that for all quantizers $\{h_l\}_{l=i}^{m}$, it holds that
\begin{align}
 \max_{l\in\{i,i+1,\dots,m-1\}}w_l(h)\geq \widehat w,\label{w>=w*_inproof}
\end{align}
where $w_l(h)$ denotes the expansion rate $w_l$ in \eqref{def,w} with the
quantization boundaries $\{h_l\}_{l=i}^{m}$.
This is done by contradiction.
Suppose that $\max_{l\in\{i,i+1,\dots,m-1\}}w_{l}(h) < \widehat w$.
Then, from \eqref{w_const}, it follows for all $l\in\{i,i+1,\dots,m-1\}$ that 
\begin{align}
 w_{l}(h)\leq \max_{l'\in\{i,i+1,\dots,m-1\}}w_{l'}(h)
 < \widehat w=w_{l}(\hat h).\label{ghb^*}
\end{align}
We shall compare $h_{l}$ with $\hat h_{l}$ for each $l$ using \eqref{ghb^*}.
For $l=i$, we have $\hat h_i=h_i$.
Substituting these into (\ref{def,w}) yields
\begin{align}
 w_i(h)&=
 \begin{cases}
  2(|a^*| + \epsilon)h_{i+1}
   \hspace{3.5em}\text{if }N\text{ is odd and }i=0,\\
  (|a^*| + \epsilon)\left(1 + \frac{\delta}{|b^*|}\right)h_{i+1}
  -(|a^*| - \epsilon)\left(1 - \frac{\delta}{|b^*|}\right)h_i\\
  \hspace{9em}\text{else},
 \end{cases}\notag\\
 w_i(\hat h)&=
 \begin{cases}
  2(|a^*|+\epsilon)\hat h_{i+1}
   \hspace{3.5em}\text{if }N\text{ is odd and }i=0,\\
  (|a^*|+\epsilon)\left(1+\frac{\delta}{|b^*|}\right)\hat h_{i+1}
  -(|a^*| - \epsilon)\left(1 - \frac{\delta}{|b^*|}\right)\hat h_i\\
  \hspace{9em}\text{else}.
 \end{cases}\notag %
\end{align}
Since $w_i(h)<w_i(\hat h)$ from (\ref{ghb^*}), it follows that
\begin{align}
 h_{i+1}<\hat h_{i+1}.\label{l=1b}
\end{align}
Moreover, by (\ref{def,w}), we have for $l=i+1,i+2\dots,m-1$ that
\begin{align}
 w_{l}(h)
 &=(|a^*| \!+\! \epsilon)\left(1 \!+\! \frac{\delta}{|b^*|}\right)h_{l+1}
  -(|a^*| \!-\! \epsilon)\left(1 \!-\! \frac{\delta}{|b^*|}\right)h_{l},\notag\\
 w_{l}(\hat h)
 &=(|a^*| \!+\! \epsilon)\left(1 \!+\! \frac{\delta}{|b^*|}\right)\hat h_{l+1}
  -(|a^*| \!-\! \epsilon)\left(1 \!-\! \frac{\delta}{|b^*|}\right)\hat h_{l}.\notag
\end{align}
By substituting these into (\ref{ghb^*}), we obtain
\begin{align}
 h_{l+1} &\leq r_ar_b h_{l}
 + \frac{\max_{l'\in\{i, i+1, \dots, m-1\}}w_{l'}(h)}
   {(|a^*|+\epsilon)\left(1+\frac{\delta}{|b^*|}\right)},\notag\\
 \hat h_{l+1} &=r_ar_b\hat h_{l}
 + \frac{\widehat w}
   {(|a^*|+\epsilon)\left(1+\frac{\delta}{|b^*|}\right)}.\notag
\end{align}
By introducing the relation (\ref{l=1b}) to the above, we recursively obtain
$h_{l}<\hat h_{l}$ for every $l\in\{i+1,i+2,\dots,m\}$.
This however is in contradiction with $\hat h_{m}=h_{m}$.
Thus \eqref{w>=w*_inproof} holds.
 
To establish \eqref{w>=w*} and \eqref{w>=w*m}, we must show the existence
of $\{\hat h_l\}_{l=i}^m$ satisfying the assumptions made at the
beginning of this proof.
First, for \eqref{w>=w*}, consider the quantizer $\{h_l^*\}_{l=0}^m$ in
\eqref{def,h*} dividing the subset $[-h_m,h_m]$ of the quantization range.
By a routine calculation and the relation $h_0=0$ in \eqref{quantizeredge}, one
can confirm that $\{h_l^*\}_{l=0}^m$ satisfies $h_0^*=h_0=0$,
$h_m^*=h_m$, and \eqref{w_const} with $i=0$ and $\widehat w=w^*_m$.
Hence, \eqref{w>=w*} follows with the equality condition.

Next, to show the case of \eqref{w>=w*m} with $[h_{m'},h_m]$, suppose that $i=m'$
and consider the following sequence for $l=0,1,\dots,m-m'$:
\begin{align}
 \hat h_{m'+l}=
 \begin{cases}
  (r_ar_b)^{l}h_{m'}+\left\{h_m-(r_ar_b)^{m-m'}h_{m'}\right\}
  \frac{1-(r_ar_b)^l}{1-(r_ar_b)^{m-m'}}\\
  \hspace{10em}\textrm{if }\epsilon>0\textrm{ or }\delta>0,\\
  h_{m'}+(h_m-h_{m'})\frac{l}{m-m'} \hspace{1.6em}\textrm{if }\epsilon=\delta=0.
 \end{cases}\notag
\end{align}
We then have that $\hat h_{m'}=h_{m'}$, $\hat h_m=h_m$, and \eqref{w_const} where
$\widehat w$ is equal to the right-hand side of \eqref{w>=w*m}.
This concludes the proof.
\end{pf}

Let us denote by $\phi_N^*$ the quantizer consisting of the boundaries
$\{h_l^*\}_{l=0}^{\lceil N/2\rceil}$ in \eqref{def,h*}.
This quantizer is optimal in the sense that it minimizes
$\max_{l\in\left\{ 0,1,\dots, \lceil N/2\rceil-1\right\}}w_l$,
and the minimum is given as $w^*_{\lceil N/2\rceil}$ from \eqref{w>=w*}.
Quantization of $\phi^*_N$ is nonuniform and becomes more so when the plant has
more uncertainty.
To see this, consider the plant with $a^*=3.0$, $\epsilon=0.5$, and $b^*=1.0$, 
and take $N=8$.
The boundaries of $\phi^*_N$ for $\delta=0.0$ are shown 
in \fref{fig,q^*}\,(a) and for $\delta=0.3$ 
in \fref{fig,q^*}\,(b).

\begin{figure}[t]
   \centering
   \includegraphics[scale=.5]{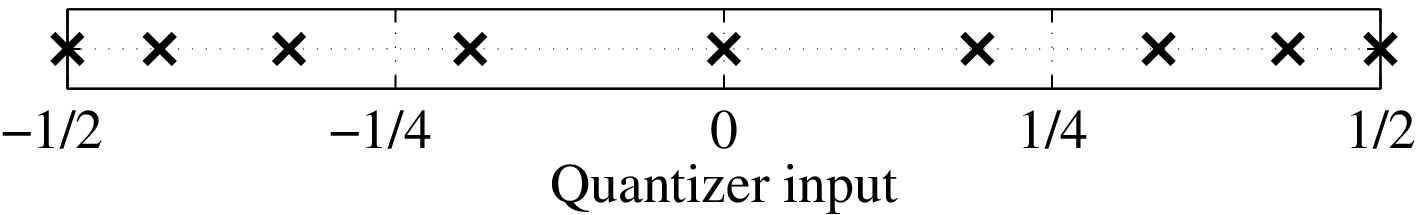}
   {\small (a)~Less uncertainty ($\delta=0.0$)}\\
   \vspace*{3mm}
   \includegraphics[scale=.5]{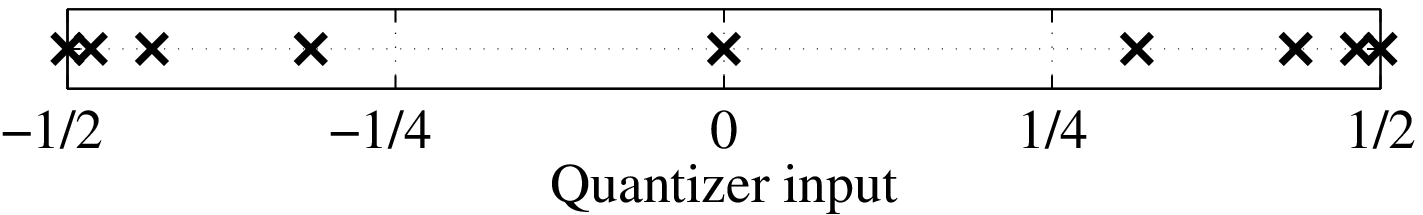}
    {\small (b)~More uncertainty ($\delta=0.3$)}\\
  \caption{Boundaries of the optimal quantizer $\phi_N^*$
  ($N=8$, $a^*=3.0$, $\epsilon=0.5$, and $b^*=1.0$)}\label{fig,q^*}
\end{figure}

The nonuniformity of $\phi^*_N$ is an outcome of minimizing the effect
of the plant uncertainty on the state estimation.
This characteristic can be explained as follows.
Due to quantization, only the interval $\calY_k$ containing
the true output $y_k$ is known to the controller.
After one time step, because of the plant instability, the interval in which
the output should be included will expand.
When the plant model is known, the expansion ratio is constant and is equal
to $|a^*|$ for any quantization cell.
However, with plant uncertainties, the ratio depends on the location of the cell.
In particular, cells further away from the origin expand more.
Moreover, to bring such cells around the origin, larger control inputs are
required compared with cells closer to the origin.
Because of the uncertainty in the input parameter $b_k$, larger control input
will result in further expansion of the interval.
This fact is illustrated in \fref{fig,uniform} when uniform quantization is 
employed.
In contrast, when the proposed quantizer $\phi^*_N$ is used, the cells
after one step have the same width (\fref{fig,optimal}).
We note that when there is no uncertainty in the plant, i.e., $\epsilon=\delta=0$,
then $\phi^*_N$ becomes the uniform quantizer.

\begin{figure}[t]
 \centering
 \begin{subfigure}{.48\linewidth}
  \centering
  \includegraphics[scale=.55]{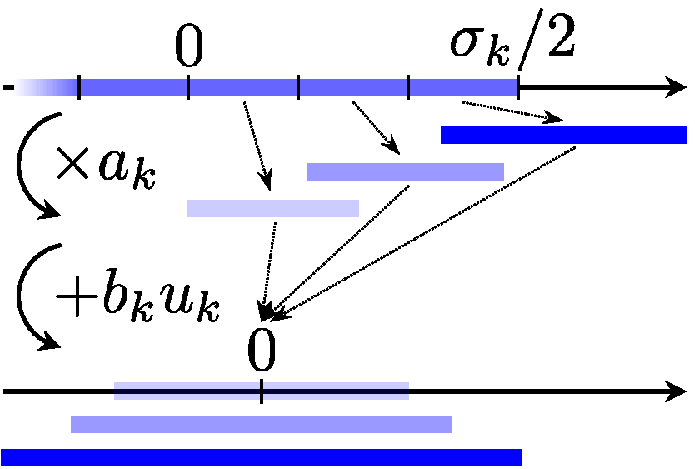}
  \caption{Uniform quantizer case}\label{fig,uniform}
 \end{subfigure}
 \begin{subfigure}{.48\linewidth}
  \centering
  \includegraphics[scale=.55]{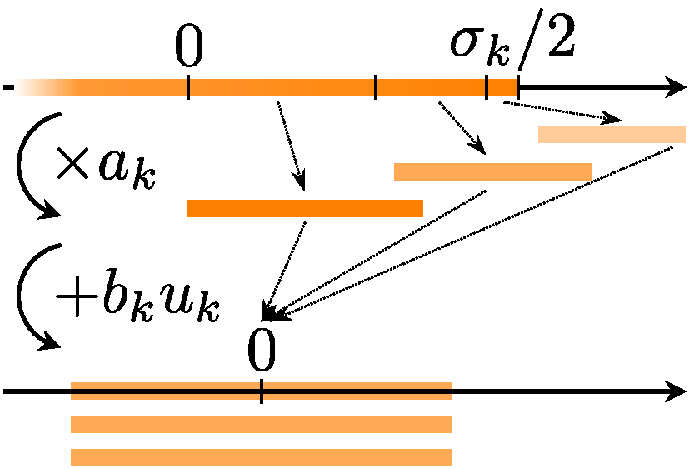}
  \caption{Optimal quantizer case}\label{fig,optimal}
 \end{subfigure}
 \caption{Expansion of quantization cells:
 With the uniform quantizer, cells further away from the origin expand more
 at each time after applying control,
 while with the optimal one, all cells result in the same width.}
\label{fig,uni-opt}
\end{figure}

The quantizer structure above is similar to that 
introduced in our previous work \cite{Okano2014a}, which studied 
plant uncertainties only in $a_{i,k}$.
We also note that nonuniform quantizers have appeared in the networked control
literature, e.g., in stabilization problems
\cite{Elia2001, Li2004} and identification problems \cite{Tsumura2009a}. 

\subsection{Derivation of the limitations in Theorem~\ref{th,scalar}}
\label{proofTh1}
We are now ready to prove Theorem~\ref{th,scalar}.
The central problem in the proof is to evaluate the expansion rate of the
estimation set $\calY_k$ during one sampling period.
First, a tight bound on the expansion rate is shown, and then
with the bound and the optimal quantizer $\phi_N^*$,
the limitations \eqref{nec,R1}--\eqref{nec,e1} for stabilizability are derived.

\begin{pf*}{Proof of Theorem~\ref{th,scalar}.}
(Necessity)
Suppose that the feedback system is stable in the sense of
Definition~\ref{def,stability} with an encoder and a controller.
We first show that $\E[\sup_{y\in\calY_k}|y|^2]\to 0$ as $k\to\infty$
implies that $\E[\sigma_k^2]\to 0$.
The estimation set $\calY_k$ available at the controller corresponds to a
quantization cell.
By \eqref{sigma_ineqb}, this set $\calY_k$ is guaranteed to contain $y_k$.
Let $d>0$ denote the smallest width of the quantization cells.
Then, from the definition of the quantizer, we have that $\mu(\calY_k)\geq d\sigma_k$,
and hence $\sup_{y\in\calY_k}|y|\geq d\sigma_k/2$.
Therefore, $\lim_{k\to\infty}\E[\sigma_k^2]=0$ follows from the mean square
stability of the system.

The rest of the proof consists of three steps.
The first step is to prove the following inequality, which provides a bound
on the expansion rate of $\sigma_k$:
\begin{align}
 \sigma_{k+1}\geq\eta_k\sigma_k. \label{lem1,sigma_eta}
\end{align}
Here, $\eta_{k}$ is the random variable defined as
\begin{align}
\eta_k:=
\begin{cases}
 |a^*|+\epsilon & \text{ if }\gamma_k=0,\\
 w_{l_k}        & \text{ if }\gamma_k=1,\\ 
\end{cases}\label{def,eta}
\end{align}
where $w_l$ is defined in (\ref{def,w}) and
$l_k\in\{0,1,\dots,\lceil N/2\rceil-1\}$ is the integer such that
$\inf_{y\in\calY_{k}}|y/\sigma_k|=h_{l_k}$, which is the index of the
quantization cell which $y_{k}$ falls into.

To establish (\ref{lem1,sigma_eta}), we recall (\ref{sigma_ineqb})
and claim that for any control input $u_k$, we have that
\begin{align}
\sigma_{k+1}
& \geq 2\!\!\!\sup_{y^-\in\calY^-_{k+1},\ b\in\calB} |y^-+bu_k|\notag\\
& \geq 2\!\!\!\sup_{y^-\in\calY^-_{k+1},\ b\in\calB} |y^-+bu^*(\calY^-_{k+1})|
  \notag\\
&= \mu(\calY^-_{k+1})+\frac{2\delta}{|b^*|}|c(\calY^-_{k+1})|
 =: \sigma_{k+1}^*, \label{lem1,lb1}
\end{align}
where the input $u^*(\cdot)$ is defined as
\begin{align}
u^*(\calY):=-\frac{c(\calY)}{b^*}\label{def,u^*}
\end{align}
for an interval $\calY$ on $\R$.
The input $u^*(\calY^-_{k+1})$ brings the midpoint of the prediction set
$\calY^-_{k+1}$ into the origin when the parameter $b_k$ is equal to $b^*$.

For the derivation of (\ref{lem1,lb1}), we first consider the case that
$c(\calY^-_{k+1})>0$ and $b^*-\delta>0$.
In this case, it is obvious that the input $u_k$ minimizing
$\sup_{y^-\in\calY^-_{k+1},\ b\in\calB}|y^-+bu_k|$ is nonpositive.
For such inputs $u_k\leq 0$, we have that
\begin{align}
 &\sup_{y^-\in\calY^-_{k+1},\ b\in\calB} |y^- + bu_k| \notag\\
 &= \max\left\{ \bY^-_{k+1}+(b^*-\delta)u_k,\ -\uY^-_{k+1}-(b^*+\delta)u_k \right\}.
 \label{supcalY^-+Bu}
\end{align}
Take an arbitrary nonpositive input and denote it as $u_k=u^*(\calY^-_{k+1})+\ud{k}$.
With this expression and (\ref{def,u^*}), the right-hand side of
(\ref{supcalY^-+Bu}) is equal to
\begin{align}
\max\Biggl\{&\frac{\mu(\calY^-_{k+1})}{2}+\frac{\delta}{b^*}c(\calY^-_{k+1})
 +(b^*-\delta)\ud{k},\notag\\
&\frac{\mu(\calY^-_{k+1})}{2}+\frac{\delta}{b^*}c(\calY^-_{k+1})
 -(b^*+\delta)\ud{k}\Biggr\},\notag
\end{align}
which takes its minimum when $\ud{k}=0$.
This proves (\ref{lem1,lb1}).
The case of $c(\calY^-_{k+1})\leq0$ or $b^*-\delta<0$
can be reduced to the above by appropriately flipping signs of $\bY_{k+1}^-$,
$\uY_{k+1}^-$, and $b^*$. 

The lower bound $\sigma_{k+1}^*$ on $\sigma_{k+1}$ in (\ref{lem1,lb1})
is in fact equal to the right-hand side of (\ref{lem1,sigma_eta}), i.e.,
\begin{align}
\sigma_{k+1}^*
=\eta_{k}\sigma_{k}.\label{lem1,muAY=eta_sigma}
\end{align}
Here, $\eta_k$ is present because the expansion rate of $\sigma_k$ is affected
by the packet losses $\gamma_k$.
To derive (\ref{lem1,muAY=eta_sigma}), we consider three cases (i)--(iii)
depending on the location of the estimation set $\calY_k$ as follows.
For simplicity, assume $a^*, b^*>0$.

(i) $\uY_{k}\geq0$:
This case occurs only when the packet arrives, i.e., $\gamma_{k}=1$.
In addition, from (\ref{def,eta}), $\eta_k=w_{l_k}$ where $N$ is even or
$l_k\neq0$.
From the basic results for products of intervals \cite{Moore2009}
and (\ref{a-e>1}), 
for the interval $\calY^-_{k+1}$, we obtain its supremum and infimum as
$\bY^-_{k+1}=(a^*+\epsilon)\bY_{k}$ and $\uY^-_{k+1}=(a^*-\epsilon)\uY_{k}$,
respectively.
Substitution of these into the left-hand side of (\ref{lem1,muAY=eta_sigma}) 
gives
\begin{align}
&\mu(\calY^-_{k+1})+\frac{2\delta}{|b^*|}|c(\calY^-_{k+1})|\notag\\
&=\left\{(a^* \!+\! \epsilon)\left(1 \!+\! \frac{\delta}{b^*}\right)h_{l_{k}+1}
   -(a^* \!-\! \epsilon)\left(1 \!-\! \frac{\delta}{b^*}\right)h_{l_{k}}\right\}\sigma_k.\notag
\end{align}
Hence, by (\ref{def,w}) the relation in (\ref{lem1,muAY=eta_sigma}) holds.

(ii) $\uY_{k}<0<\bY_{k}$:
In this case, we have
\begin{align}
\bY^-_{k+1}=(a^*+\epsilon)\bY_{k},\quad
 \uY^-_{k+1}=(a^*+\epsilon)\uY_{k}.\label{lem1,bAY,uAY_2}
\end{align}
Moreover, because of the symmetry in the quantization cells, it holds that
$c(\calY^-_{k+1})=0$ and hence $u^*(\calY^-_{k+1})=0$.
Thus, $\sigma^*_{k+1}=\mu(\calY^-_{k+1})$.
To compute this width $\mu(\calY^-_{k+1})$, consider the following two cases:
(ii-1) If $\gamma_{k}=0$, then $\calY_{k}=[-\sigma_{k}/2,\sigma_{k}/2]$.
Hence, by (\ref{lem1,bAY,uAY_2}), we have
$\mu(\calY^-_{k+1})=(a^*+\epsilon)\sigma_{k}$.
(ii-2) Otherwise, $N$ must be odd and $l_{k}=0$ from the symmetry of the quantizer
and the condition (ii). Thus, 
$\mu(\calY^-_{k+1})=2(a^*+\epsilon)h_1\sigma_{k}$. 
Hence, (\ref{lem1,muAY=eta_sigma}) holds for this case also.

(iii) $\bY_{k}\leq 0$: This case can be reduced to (i).

From (\ref{lem1,lb1}) and (\ref{lem1,muAY=eta_sigma}),
we have established (\ref{lem1,sigma_eta}).

As the second step, we consider the mean squares of both sides of
(\ref{lem1,sigma_eta}) and derive an inequality regarding the expansion rate of
$\sigma_k$.
Note that due to packet losses, the expansion rate $\eta_k$ is a random variable
that depends on the previous channel state.
This time dependency causes difficulties in the analysis.
To avoid this, we consider intervals between the times at which
packets successfully arrive at the controller side;
it is known that the intervals become an i.i.d.\ process \cite{Xie2009}.
We formally state this fact in the following.

Let $t_j$, $j\in\Zp$, be the times at which packets arrive,
i.e., $\gamma_{t_j}=1$.
From the assumption on the initial communication, stated in Section \ref{sec,problem},
we have $0=t_0<t_1<\cdots$.
Then, denote the interval between the times $t_{j-1}$ and $t_j$ by
$\tau_j:=t_{j}-t_{j-1},\ j\geq1$.
The process $\{\tau_j\}_j$ is i.i.d.\ and, for each $j$, it holds that
\begin{align}
\Prob(\tau_j=i)=\begin{cases}
		  1-\pl & \text{if }i=1,\\
		  \pl\pr(1-\pr)^{i-2} & \text{if } i>1.
		 \end{cases}\label{sojourntime}
\end{align}
From (\ref{lem1,sigma_eta}) and the fact that
\begin{align}
\gamma_k=\begin{cases}
	   1 & \text{if }k=t_j,\\
	   0 & \text{if }k\in[t_j+1,t_{j+1}),
	  \end{cases}\notag
\end{align}
a lower bound on the expansion of $\sigma_k$ from time $t_j$ to $t_{j+1}$ is
given by
$\sigma_{t_{j+1}}\geq (|a^*|+\epsilon)^{\tau_{j+1}-1}w_{l_{t_j}}\sigma_{t_j}$.
Here, the value of $w_{l_{t_j}}$ varies depending on $l_{t_j}$, which
is the index of the cell that $y_{t_j}$ fell in.
Since we have to take account of all possible deterministic perturbations,
we consider the maximum over the suffix of $w_{l_{t_j}}$:
\begin{align}
 \sigma_{t_{j+1}}
 \geq
 (|a^*|+\epsilon)^{\tau_{j+1}-1}
 \max_{l\in L_{t_j}}w_l \sigma_{t_j},\label{sigma_maxw_Lk}
\end{align}
where $L_{t_j}$ contains the set of all possible $l_{t_j}$,
which is a subset of the index set $\{0,1,\dots,\lceil N/2\rceil-1\}$.
More precisely, at time $k$, $L_k$ is the indices of the quantization cells
which intersect with $\{y^-+bu_{k-1} : y^-\in\calY^-_k, b\in\calB\}$.

As to the right-hand side of \eqref{sigma_maxw_Lk}, we shall prove that
\begin{align}
 \max_{l\in L_k}w_l\sigma_k
 \geq w^*_{\ceil{N/2}} \sigma_k^*,
 \label{wsigma>=w*sigma*}
\end{align}
where $w^*_\bullet$ and $\sigma_k^*$ are given by \eqref{def,w*} and
\eqref{lem1,lb1}, respectively.
Let $\hcalY_{k}^-$ be the union of the quantization cells which have intersection
with $\{y^-+bu_{k-1} : y^-\in\calY^-_k, b\in\calB\}$.
Consider the following two cases.

(i) $\inf\hcalY_{k}^-\leq 0 \leq \sup\hcalY_{k}^-$:
Let $m\in\{1,2,\dots,\ceil{N/2}\}$ be the cardinality of $L_k$.
We then have that $\max_{l\in L_k}w_l=\max_{l\in\{0,1,\dots,m-1\}}w_l$ and
$\sup_{y^-\in\hcalY_k^-}|y^-|=h_{m}\sigma_k$.
Thus, from \eqref{w>=w*} in Lemma~\ref{lem,LBofw}, it holds that
\begin{align}
 \max_{l\in L_k}w_l\sigma_k
 &\geq \frac{w^*_m}{h_m}\sup_{y^-\in\hcalY_k^-}|y^-|
 \geq 2 w^*_{\ceil{N/2}} \sup_{y^-\in\hcalY_k^-}|y^-|,\notag
\end{align}
where we have used $w^*_m/h_m\geq 2w^*_{\ceil{N/2}}$
to obtain the second inequality.
Furthermore, since $\hcalY_k^-\supseteq \{y^-+bu_{k-1} : y^-\in\calY_k^-,\ b\in\calB\}$
and by \eqref{lem1,lb1}, we have $2\sup_{y^-\in\hcalY_k^-}|y^-|\geq \sigma_k^*$.
Thus, we obtain \eqref{wsigma>=w*sigma*} for this case.

(ii) $\inf\hcalY_{k}^->0$ or $\sup\hcalY_{k}^-<0$:
In this case, we can define integers $m'$ and $m$ such that
$1\leq m'<m\leq\ceil{N/2}$, $\inf_{y^-\in\hcalY_{k}^-}|y^-|=h_{m'}\sigma_k$,
and $\sup_{y^-\in\hcalY_{k}^-}|y^-|=h_{m}\sigma_k$.
By using \eqref{w>=w*m} in Lemma~\ref{lem,LBofw}, it follows that
\begin{align}
 &\max_{l\in L_k}w_l\sigma_k\notag\\
 &\geq
 \begin{cases}
  \left\{\sup_{y^-\in\hcalY_{k}^-}|y^-|\!
   -\!(r_ar_b)^{m-m'}\inf_{y^-\in\hcalY_{k}^-}|y^-| \right\}\\
  \times\left( |a^*|\!+\! \epsilon \right)\left( 1\!+\!\frac{\delta}{|b^*|} \right)
   \!\frac{1-r_ar_b}{1-(r_ar_b)^{m-m'}}
   \hspace{1em} \text{if } \epsilon\!>\!0 \text{ or }\delta\!>\!0,\\
  \mu(\hcalY_k^-)
   \frac{|a^*|}{m-m'}
   \hspace{7.8em} \text{if }\epsilon=\delta=0.
 \end{cases}
 \label{LB_wl_ii}
\end{align}
If $\epsilon=\delta=0$, noticing that $\mu(\hcalY^-_k)\geq\sigma_k^*$, we have
\eqref{wsigma>=w*sigma*} for this case.
When $\epsilon>0$ or $\delta>0$, a routine calculation shows that
\begin{align}
 &\left\{\sup_{y^-\in\hcalY_{k}^-}|y^-|
   -(r_ar_b)^{m-m'}\inf_{y^-\in\hcalY_{k}^-}|y^-| \right\}
 \frac{1}{1-(r_ar_b)^{m-m'}} \notag\\
 &\geq \left(\frac{\mu(\hcalY_k^-)}{2}+\frac{\delta}{|b^*|}
 \left|c(\hcalY_k^-)\right|\right) \frac{1}{1-t(r_ar_b)^{m}}\notag\\
 &\geq\frac{\sigma_k^*}{2\{1-t(r_ar_b)^{m}\}}.\label{LB_wl_iia}
\end{align}
Here, for the first inequality, we have used
$\sup_{y^-\in\hcalY_{k}^-}|y^-|=\mu(\hcalY_k^-)/2+|c(\hcalY_k^-)|$,
$\inf_{y^-\in\hcalY_{k}^-}|y^-|=-\mu(\hcalY_k^-)/2+|c(\hcalY_k^-)|$, and
$0<tr<1$, and the second one follows by $\hcalY_k^-\supseteq\calY_k^-$.
From \eqref{LB_wl_ii} and \eqref{LB_wl_iia}, and by $m\leq\ceil{N/2}$,
we arrive at \eqref{wsigma>=w*sigma*}.

From \eqref{lem1,lb1}, \eqref{sigma_maxw_Lk}, and \eqref{wsigma>=w*sigma*},
we have
\begin{align}
 \E[{\sigma_{t_{j+1}}^*}^2]
 \geq \E[(|a^*|+\epsilon)^{2(\tau_{j+1}-1)}]
 \left({w^*_{\ceil{N/2}}}\right)^2\E[{\sigma_{t_j}^*}^2].
 \label{MSsigma_ineq}
\end{align}
Notice that $\tau_{j+1}$ is independent of $\sigma_{t_j}$.
Since $\E[\sigma_k^2]\to 0$ and by \eqref{lem1,lb1},
the coefficient of $\E[{\sigma_{t_j}^*}^2]$ in \eqref{MSsigma_ineq} is
smaller than $1$.
After some computation using \eqref{sojourntime} and that $\{\tau_j\}_j$
is i.i.d., we obtain
\begin{align}
 \E[(|a^*|+\epsilon)^{2(\tau_{j+1}-1)}]
 =1+\frac{p\left\{(|a^*|+\epsilon)^2-1\right\}}
    {1-(1-q)(|a^*|+\epsilon)^2}=\nu^2\notag
\end{align}
and $|(1-q)(|a^*|+\epsilon)^2|<1$.
The inequality boils down to
\begin{align}
 0<(1-q)(|a^*|+\epsilon)^2<1,\label{expcconvcond}
\end{align}
which implies $\nu^2>0$.
Hence, we arrive at
\begin{align}
 \nu w^*_{\ceil{N/2}} <1. \label{nec_nuw}
\end{align}

The last step is to derive (\ref{nec,R1})--(\ref{nec,e1}) from \eqref{expcconvcond}
and \eqref{nec_nuw}.
First, suppose that $\epsilon$ or $\delta$ is positive and 
that $N$ is even.
Then, with the definition (\ref{def,w*}), it follows that
$w^*_{\ceil{N/2}}={\Delta}/\{1-(r_ar_b)^{\lceil N/2\rceil}\}$.
Substituting this into \eqref{nec_nuw}, we obtain $(r_ar_b)^{N/2}<1-\nu\Delta$
and by taking logarithm of both sides,
\begin{align}
 N>\Nnece:=2\frac{\log(1-\Delta\nu)}{\log (r_a r_b)}.\notag
\end{align}
Note that
\begin{align}
 1-\nu\Delta>0\label{nuDelta<1}
\end{align}
since $(r_ar_b)^{N/2}>0$.
If $N$ is odd, then similarly, from \eqref{nec_nuw}, we have that
\begin{align}
 N>\Nneco:=2\frac{\log(1-\Delta\nu)-\log t}{\log (r_a r_b)}-1,\notag
\end{align}
where $t:=\left(1+\delta/|b^*|\right) / \left(1-\epsilon/|a^*|\right)$.
Comparing the lower bounds $\Nnece$ and $\Nneco$, by the assumption
(\ref{b-d>0}) on $b^*$ and (\ref{a-e>1}), 
we can show $\Nneco>\Nnece$.
Thus, the smaller bound $\Nnece$ appears as the data rate limitation in
(\ref{nec,R1}).
To establish (\ref{nec,p1}) and (\ref{nec,e1}), we see
from \eqref{expcconvcond} and \eqref{nuDelta<1} that
\begin{align}
 (1-\Delta^2)\left\{1-(1-q)(|a^*|+\epsilon)^2\right\}
  > \Delta^2\left\{(|a^*|+\epsilon)^2-1\right\}.\notag
\end{align}
Noticing that the right-hand side is positive and so is the left-hand side,
we have (\ref{nec,p1}) and (\ref{nec,e1}) after some calculations.

If $\epsilon=\delta=0$,
we have that $w^*_{\ceil{N/2}}={|a^*|}/{N}$ by \eqref{def,w*}.
Noticing this relation and applying the same analysis, with \eqref{nec_nuw},
we have (\ref{nec,R1})--(\ref{nec,e1}) for this case also.

(Sufficiency)
We employ the control scheme where $\sigma_{k+1}$ is determined so that
the equality holds in (\ref{sigma_ineqb}) and $u_k=u^*(\calY^-_{k+1})$.
With this control, we have the equality in (\ref{lem1,sigma_eta}).
Moreover, let the quantizer be the optimal one $\phi^*_N$, where the boundaries
are $\{h_l^*\}_{l=0}^{\ceil{N/2}}$ in \eqref{def,h*}.
Then, equality holds in \eqref{wsigma>=w*sigma*} also.
From (\ref{nec,p1}) and (\ref{nec,e1}), we have that $\nu^2>0$ and \eqref{expcconvcond},
and then \eqref{nuDelta<1}.
In addition, with \eqref{nec,R1}, the key inequality \eqref{nec_nuw} in the necessity
part follows and thus, $\E[\sigma_k^2]\to 0$.
Since $\sigma_k/2\geq\sup_{y_\in\calY_k}|y|$ by definition of the quantizer,
we have that $\E[\sup_{y\in\calY_k}|y|^2]\to 0$ as $k\to\infty$.
\end{pf*}

\section{General order plants case}\label{sec,multi}
In this section, we consider general order plants in (\ref{AR}),
and present a control scheme to make the feedback system MSS
along with a sufficient condition providing a stability test. 
In the course, we will see that results of Markov jump
linear systems \cite{Costa2005} play a central role.
A related approach can be found in \cite{Minero2013}, where the
known plants case has been studied. 

Given a data rate $R=\log N$ and a quantizer $\{h_l\}_{l=0}^{\ceil{N/2}}$,
we set the control law as follows:
In the encoder and the decoder, the scaling parameter $\sigma_k$ is determined by
\begin{align}
 \sigma_k=\mu(\calY^-_{k})+\frac{2\delta}{|b^*|}|c(\calY^-_{k})|\label{suf,sigma}
\end{align}
at each time $k$.
Here $c(\cdot)$ is the midpoint of an interval.
Furthermore, the control input $u_k$ is given as
\begin{align}
 u_k=-\frac{c(\calY^-_{k+1})}{b^*}.\label{suf,u}
\end{align}

Next, we introduce some notation.
For $i=1,2,\dots,n$, let the random variables $\theta_{i,k}$ be given by
\begin{align}
 \theta_{i,k}:=
 \begin{cases}
  |a_i^*|+\epsilon_i & \text{if }\gamma_{k-i+1}=0,\\
  \bw_{i} & \text{if }\gamma_{k-i+1}=1.
 \end{cases}\label{def,thetab}
\end{align}
Here, $\bw_i$ is defined for the given quantizer as follows:
\begin{align}
 \bw_i &\!:=\!
 \begin{cases}
  \max\{\bwzero_i,\bwone_i\}
  & \text{if }N \text{ is odd and } \calA_i\not\ni 0,\\
  \max\left\{\epsilon_i+\delta\frac{|a^*_i|}{|b^*|},\ \bwzero_i\right\}
  & \text{if }N \text{ is odd and } \calA_i\ni 0,\\
  \bwone_i
  & \text{if }N \text{ is even and } \calA_i\not\ni 0,\\
  \epsilon_i+\delta\frac{|a^*_i|}{|b^*|}
  & \text{if }N \text{ is even and } \calA_i\ni 0,
 \end{cases}\notag
\end{align}
where $\bwzero_i$ and $\bwone_i$ are given by
\begin{align}
 \bwzero_i &:= 2(|a_i^*|+\epsilon_i) h_1, \label{def,w0bar}\\
 \bwone_i  &:= \max_{l\in\{0,\dots,\lceil N/2\rceil-1\}}
 \Biggl\{(|a_i^*|+\epsilon_i)\left(1+\delta\frac{|a_i^*|}{|b^*|}\right)h_{l+1}\notag\\
  &\hspace{6.5em}
  -(|a_i^*|-\epsilon_i)\left(1-\delta\frac{|a_i^*|}{|b^*|}\right)h_l\Biggr\}.
  \label{def,w1bar}
\end{align}
We will see in the proof later that $\bw_i$ is used to express an upper
bound of the expansion rate of the quantization cells enlarged by the parameter
$a_{i,k}$.
Moreover, let the matrix $H_{\Gamma_k}$ containing
$\theta_{1,k},\dots,\theta_{n,k}$ be
\begin{align}
 H_{\Gamma_k}:=\left[\begin{array}{cccc}
   0& 1& \cdots& 0\\
   \vdots& \ddots& \ddots& \vdots\\
   0& 0& \cdots& 1\\
   \theta_{n,k} & \theta_{n-1,k} & \cdots&\theta_{1,k} 
 \end{array}\right]. \label{def,H}
\end{align}
Here, $\Gamma_k$ is the random vector defined as
$\Gamma_k:=[\gamma_{k-n+1}\ \gamma_{k-n+2}$ $\cdots\ \gamma_{k}]$,
which is determined by the results of the past $n$ communications.
The transition probability matrix $P\in\R^{2^n\times 2^n}$ for the 
random process $\{\Gamma_k\}_k$ is given by, for $n\geq2$,
\begin{align}
  P:=
 \left[\begin{array}{c}
  I_{2^{n-2}}\\
  I_{2^{n-2}}
 \end{array}\right]\otimes Q,\quad
  Q:=\left[\begin{array}{cccc}
     1-\pr& \pr& 0& 0  \\
       0& 0& \pl& 1-\pl
\end{array}\right],\notag
\end{align}
where $\otimes$ is the Kronecker product, and when $n=1$,
\begin{align}
 P:=\left[\begin{array}{cc}
     1-\pr& \pr\\
     \pl  & 1-\pl
 \end{array}\right].\notag
\end{align}
Finally, we define the matrix $F$ using $H_{\Gamma_k}$ and $P$ by
\begin{align}
 &F:=F_1F_2,\label{def,F}
\end{align}
where $F_1:=P^T \otimes I_{n^2}$, 
$F_2:=\diag(H_{\Gamma^{(1)}}\otimes H_{\Gamma^{(1)}},\dots,
 H_{\Gamma^{(2^n)}}\otimes H_{\Gamma^{(2^n)}})$, and
$\Gamma^{(1)},\dots,\Gamma^{(2^n)}$ represent all possible 
vectors of $\Gamma_k$ indexed arbitrarily.
Let $\rho(\cdot)$ be the spectral radius of a matrix.

We are ready to present the main result of the subsection.
\begin{thm}\label{th,suf}
Given the data rate $R=\log N$, the loss probability $p\in[0,1)$, and
the quantizer $\{h_l\}_{l=0}^{\ceil{N/2}}$, if the matrix $F$ in (\ref{def,F}) satisfies
\begin{align}
 \rho(F)<1\label{suf_cond},
\end{align}
then under the control law using (\ref{suf,sigma}) and (\ref{suf,u}),
the feedback system
is MSS.
\end{thm}

For the scalar plants case, the inequality \eqref{suf_cond} holds
if and only if $\nu \bw_1<1$ and $\left\{(1-q)(|a^*|+\epsilon)^2+(1-p)\bw_1^2\right\}/2\leq 1$.
Notice that the first inequality is equal to \eqref{nec_nuw} in the proof of
Theorem~\ref{th,scalar} when the quantizer is the optimal one $\phi_N^*$.
Thus, for this case, the condition \eqref{suf_cond} is tight in the sense
that if $R\in\N$, $p,q\in(0,1)$, and $\Delta$ satisfy \eqref{nec,R1}--\eqref{nec,e1},
then there always exist an encoder and a controller such that \eqref{suf_cond} holds.

To establish the theorem, consider the Markov jump system
\begin{align}
 z_{k+1}=H_{\Gamma_k}z_k,\quad z_0:=[\sigma_{-n+1}\ \sigma_{-n}\ \cdots\
  \sigma_{0}]^T.
 \label{MJLS-z}
\end{align}
The stability of the feedback system can be reduced to
that of this system. This is shown in the following lemma,
whose proof is given in the Appendix.

\begin{lem}\label{lem,for_Th_suf}
If the Markov jump system (\ref{MJLS-z}) is stable in the sense that
$\E[z_kz_k^T]$ converges to the zero matrix,
the original feedback system is MSS with the control law (\ref{suf,sigma}) and
(\ref{suf,u}).
\end{lem}

\begin{rem}
In \cite{Okano2014a}, a necessary condition for the multidimensional plants case,
which is similar to the scalar case result, is given.
This result has been shown under the setup that the structures of state estimators
and controllers are constrained as \eqref{def,calY^-} and \eqref{controller}.
As we discussed in Remark~\ref{rem,structures}, this may cause some conservativeness.
Therefore, in this paper, we do not present the result corresponding to
Theorem~\ref{th,scalar}.
\end{rem}

\section{Conclusion}\label{sec,conclusion}
In this paper, we have addressed a stabilization problem of parametrically 
uncertain plants 
over data rate limited channels subject to random data losses.
The result for the scalar case establishes limitations and trade-off
relationships for stability among the data rate, the transition probabilities
of the channel states, and the uncertainty bounds.
As mentioned in the Introduction, uncertain systems in networked settings have
not been studied much.
We plan to extend our research in this area in the future. 

\smallskip
\textit{Acknowledgment}: The authors would like to thank
M.~Fujita, S.~Hara, and R.~Tempo for helpful discussions on this work.
We are also grateful to the anonymous reviewers for their comments that
helped us improve the paper quality.

\appendix

\section{Proof of Theorem \ref{th,suf}}\label{sec,pf_Th_suf}

\begin{pf*}{Proof of Lemma \ref{lem,for_Th_suf}}
We first verify that the mean square stability of $\{\sigma_k\}_k$ implies that
the feedback system is MSS under the control law (\ref{suf,sigma}) and (\ref{suf,u}).
This is done by substituting (\ref{suf,u}) into (\ref{AR}) and by
the definition (\ref{def,calY^-}) of $\calY_{k+1}$ to obtain
\begin{align}
 |y_{k+1}|
 &=\left|a_{1,k}y_k+\cdots+a_{n,k}y_{k-n+1}
  -b_k\frac{c(\calY^-_{k+1})}{b^*}\right|\notag\\
 &\leq\frac{1}{2}\left(\mu(\calY^-_{k+1})+\frac{2\delta}{|b^*|}|c(\calY^-_{k+1})|\right)
  =\frac{\sigma_{k+1}}{2}.\notag
\end{align}

Next, to establish that the stability of (\ref{MJLS-z}) implies that $\{\sigma_k\}_k$
is MSS, we prove the following relation:
\begin{align}
 \sigma_k\leq(z_k)_n \text{ for }k=0,1,\dots,\label{sigma-z}
 \end{align}
where $(\cdot)_n$ is the $n$th element of a vector.
The scaling parameter (\ref{suf,sigma}) can be decomposed as
\begin{align}
 \sigma_{k+1}
 &=\sum_{i=1}^n\mu(\calA_{i}\calY_{k-i+1})
  +\frac{2\delta}{|b^*|}|c(\calY^-_{k+1})|\notag\\
 &\leq\sum_{i=1}^n \left\{\mu(\calA_{i}\calY_{k-i+1})
  +\frac{2\delta}{|b^*|}|c(\calA_i\calY_{k-i+1})|\right\}.\label{sigmadecomposition}
\end{align}
Here, the equality follows from the Brunn-Minkowski theorem %
and the inequality 
from applying the triangle inequality to the second term.
Next, we explicitly compute the width $\mu(\calA_i\calY_{k-i+1})$ of the product
of intervals $\calA_i$ and $\calY_{k-i+1}$.
Recall $\calA_i=[a_i^*-\epsilon_i,a_i^*+\epsilon_i]$.
Based on basic results for interval products \cite{Moore2009}, we can obtain
\begin{align}
 \mu\left(\calA_i\calY_{k-i+1}\right)
 \!=\!
 \begin{cases}
   \left(|a_i^*|+\epsilon_i\right)\mu(\calY_{k-i+1})\\
    \hspace{3em} \text{if }\calY_{k-i+1}\ni 0,\\
   |a_i^*|\mu(\calY_{k-i+1}) \!+\! \epsilon_i|\bY_{k-i+1}\!+\!\uY_{k-i+1}|\\
    \hspace{3em}  \text{if }\calY_{k-i+1}\not\ni 0 \text{ and } \calA_{i}\not\ni 0,\\
   2\epsilon_i\max\left\{|\bY_{k-i+1}|,\,|\uY_{k-i+1}|\right\}\\
    \hspace{3em}  \text{if }\calY_{k-i+1}\not\ni 0 \text{ and } \calA_{i}\ni 0,
 \end{cases}%
\label{muAYb}
\end{align}
for $i=1,2,\dots,n$.
Similarly, the absolute value of the center of the interval $\calA_i\calY_{k-i+1}$
can be computed as
\begin{align}
 &|c(\calA_i\calY_{k-i+1})|\notag\\
 &=\begin{cases}
   0 \hspace{6.75em} \text{if }\calY_{k-i+1}\ni0,\\
   \frac{1}{2}\biggl[(|a_i^*|+\epsilon_i)\max\left\{|\bY_{k-i+1}|,\ |\uY_{k-i+1}|\right\}\\
   \hspace{1em} + (|a_i^*|-\epsilon_i)\min\left\{|\bY_{k-i+1}|,\ |\uY_{k-i+1}|\right\}\biggr]\\
   \hspace{7.2em} \text{if }\calY_{k-i+1}\not\ni 0 \text{ and } \calA_{i}\not\ni 0,\\
   |a_i^*|\max\left\{|\bY_{k-i+1}|,\ |\uY_{k-i+1}|\right\}\\
   \hspace{7.2em} \text{if }\calY_{k-i+1}\not\ni 0 \text{ and } \calA_{i}\ni 0.
  \end{cases}\label{cAYb}
\end{align}
We use (\ref{muAYb}) and (\ref{cAYb}) to obtain an upper bound on the
$i$th term in (\ref{sigmadecomposition}) over all possible $\calY_{k-i+1}$ as
\begin{align}
 \mu(\calA_{i}\calY_{k-i+1}) + \frac{2\delta}{|b^*|}|c(\calA_i\calY_{k-i+1})|
 \leq\theta_{i,k}\sigma_{k-i+1}.\label{muAY_thetasigma}
 \end{align}
This is shown by examining the following three cases depending on the packet
loss process $\gamma_{k-i+1}$ and $N$.

(i) $\gamma_{k-i+1}=0$:
In this case, by construction in Section~\ref{sec,problem},
$\calY_{k-i+1}$ is the entire input range, that is,
$\calY_{k-i+1}=[-\sigma_{k-i+1}/2,\sigma_{k-i+1}/2]$.
This interval contains the origin as an interior point.
Thus, by (\ref{muAYb}) and (\ref{cAYb}), we have
\begin{align}
 \mu\left(\calA_i\calY_{k-i+1}\right)+\frac{2\delta}{|b^*|}|c(\calA_i\calY_{k-i+1})|
  =(|a_i^*|+\epsilon_i)\sigma_{k-i+1.\notag}
\end{align}
From (\ref{def,thetab}), we have $|a_i^*|+\epsilon_i=\theta_{i,k}$
for this case and hence, (\ref{muAY_thetasigma}) holds.

(ii) $\gamma_{k-i+1}=1$ and $N$ is odd:
We must deal with two cases.
(ii-1) $\uY_{k-i+1}<0<\bY_{k-i+1}$: %
This implies that $\calY_{k-i+1}$ is the center quantization cell,
i.e., its boundaries $(\uY_{k-i+1},\bY_{k-i+1})$ are
$(-h_1\sigma_{k-i+1},h_1\sigma_{k-i+1})$ (see (\ref{quantizeredge}) for the
definition of the cells).
Thus, from (\ref{muAYb}) and (\ref{cAYb}), we have
\begin{align}
 \mu\left(\calA_i\calY_{k-i+1}\right)+\frac{2\delta}{|b^*|}|c(\calA_i\calY_{k-i+1})|
 =\bwzero_i\sigma_{k-i+1},\label{suf,muAY,ii-1b}
\end{align}
where $\bwzero_i$ is defined in (\ref{def,w0bar}).

(ii-2) Otherwise: %
For this case, $\calY_{k-i+1}$ does not contain the origin as an interior
point and hence, by (\ref{quantizeredge}), its boundaries $(\uY_{k-i+1},\bY_{k-i+1})$
are equal to $(h_l\sigma_{k-i+1},h_{l+1}\sigma_{k-i+1})$ or
$(-h_{l+1}\sigma_{k-i+1},-h_{l}\sigma_{k-i+1})$ for some index $l$.
Therefore, with (\ref{muAYb}) and (\ref{cAYb}),
\begin{align}
 &\mu(\calA_i\calY_{k-i+1})+\frac{2\delta}{|b^*|}|c(\calA_i\calY_{k-i+1})|\notag\\
 &=\begin{cases}
    \left\{(|a_i^*|+\epsilon_i)\left(1+\frac{\delta}{|b^*|}\right)h_{l+1}
    -(|a_i^*|-\epsilon_i)\left(1+\frac{\delta}{|b^*|}\right)h_l\right\}\sigma_{k-i+1}\\
    \hspace{16em}\text{if }\calA_{k-i+1}\not\ni 0,\\
    2\left(\epsilon_i+\delta\frac{|a_i^*|}{|b^*|}\right)h_{l+1}\sigma_{k-i+1}
    \hspace{7.1em} \text{if }\calA_{k-i+1}\ni 0.
 \end{cases}\notag
\end{align}
Taking the maximum of the right-hand side of the above equality over $l\in\{1,2,\dots,\lceil N/2\rceil-1\}$,
we have
\begin{align}
 &\max_l\mu\left(\calA_i\calY_{k-i+1}\right)+\frac{2\delta}{|b^*|}|c(\calA_i\calY_{k-i+1})|\notag\\
 &=\begin{cases}
   \bwone_i\sigma_{k-i+1} & \text{if }\calA_{k-i+1}\not\ni 0,\\
   \left(\epsilon_i+\delta\frac{|a_i^*|}{|b^*|}\right)\sigma_{k-i+1}  & \text{if }\calA_{k-i+1}\ni 0,
  \end{cases}\label{suf,muAY,ii-2b}
\end{align}
where $\bwone_i$ is defined in (\ref{def,w1bar}).
From (\ref{suf,muAY,ii-1b}) and (\ref{suf,muAY,ii-2b}), we confirm
(\ref{muAY_thetasigma}) for the case (ii) also.

(iii) $\gamma_{k-i+1}=1$ and $N$ is even:
This case can be reduced to (ii-2) since $\calY_{k-i+1}\not\ni0$ holds.

From (\ref{sigmadecomposition}) and (\ref{muAY_thetasigma}), we have
$\sigma_{k+1}\leq\sum^{n}_{i=1}\theta_{i,k}\sigma_{k-i+1}$.
Here, notice that the right-hand side is equal to
the $n$th entry of the vector
$H_{\Gamma_k}[\sigma_{k-n+1}\ \sigma_{k-n+2}\ \cdots\ \sigma_{k}]^T$,
where the matrix $H_{\Gamma_k}$ is given in (\ref{def,H}).
Thus, using the $n$th state of the Markov jump system (\ref{MJLS-z}),
we obtain (\ref{sigma-z}).
This implies that if $\E[z_kz_k^T]$ goes to the zero matrix as $k\to\infty$,
then $\E[\sigma_k^2]\to0$.
\end{pf*}

From the stability result \cite[Theorem 3.9]{Costa2005} for Markov jump systems,
the inequality (\ref{suf_cond}) is equivalent to that (\ref{MJLS-z}) is MSS.
Therefore, from Lemma~\ref{lem,for_Th_suf}, this inequality (\ref{suf_cond})
is a sufficient condition for mean square stability of the original feedback system.
This concludes the proof of Theorem~\ref{th,suf}.


\begin{thebibliography}{10}
\bibitem{Tatikonda2004}
S.~Tatikonda and S.~Mitter, ``{Control under communication constraints},''
  \emph{\IEEEJAC}, 49: 1056--1068, 2004.

\bibitem{Nair2004}
G.~N. Nair and R.~J. Evans, ``{Stabilizability of stochastic linear systems
  with finite feedback data rates},'' \emph{\SIAMCO}, 43: 413--436, 2004.

\bibitem{Nair2004a}
G.~N. Nair, R.~J. Evans, I.~M.~Y. Mareels, and W.~Moran,
``Topological feedback entropy and nonlinear stabilization,'' \emph{\IEEEJAC},
49: 1585--1597, 2004.

\bibitem{Savkin2006}
A.~V.~Savkin, ``Analysis and synthesis of networked control systems: Topological
entropy, observability, robustness and optimal control,'' \emph{Automatica},
42: 51--62, 2006.

\bibitem{Colonius2013}
F.~Colonius, C.~Kawan, and G.~Nair, ``A note on topological feedback entropy and
invariance entropy,'' \emph{\SysCL}, 62: 377--381, 2013.

\bibitem{Nair2007}
G.~N. Nair, F.~Fagnani, S.~Zampieri, and R.~J. Evans, ``{Feedback control under
  data rate constraints: An overview},'' \emph{\IEEEJPROC}, 95: 108--137, 2007.

\bibitem{Heemels2010}%
W.~P.~M.~H.~Heemels, A.~R.~Teel, N.~van de Wouw, and D.~Nesic,
``Networked control systems with communication constraints: Tradeoffs
between transmission intervals, delays and performance,'' 
\emph{\IEEEJAC}, 55: 1781--1796, 2010.

\bibitem{Yuksel2014}%
S. Y\"uksel, ``Jointly optimal LQG quantization and control policies for
multi-dimensional systems,'' 
\emph{\IEEEJAC}, 59: 1612--1617, 2014.

\bibitem{Hespanha2007}
J.~P. Hespanha, P.~Naghshtabrizi, and Y.~Xu, ``{A survey of recent results in
  networked control systems},'' \emph{\IEEEJPROC}, 95: 138--172, 2007.

\bibitem{Schenato2007}
L.~Schenato, B.~Sinopoli, M.~Franceschetti, K.~Poolla, and S.~S.~Sastry,
  ``{Foundations of control and estimation over lossy networks},''
  \emph{\IEEEJPROC}, 95: 163--187, 2007.

\bibitem{You2011a}
K.~You and L.~Xie, ``{Minimum data rate for mean square stabilizability of
  linear systems with Markovian packet losses},'' \emph{\IEEEJAC},
  56: 772--785, 2011.

\bibitem{Minero2013}
P.~Minero, L.~Coviello, and M.~Franceschetti, ``{Stabilization over Markov
  feedback channels: The general case},'' \emph{\IEEEJAC}, 58: 349--362, 2013.

\bibitem{Phat2004}
V.~N. Phat, J.~Jiang, A.\,V. Savkin, and I.~R. Petersen, ``{Robust stabilization
  of linear uncertain discrete-time systems via a limited capacity
  communication channel},'' \emph{\SysCL}, 53: 347--360, 2004.

\bibitem{Fu2010}
M.~Fu and L.~Xie, ``{Quantized feedback control for linear uncertain
  systems},'' \emph{\IJRN}, 20: 843--857, 2010.

\bibitem{Martins2006}
N.~C. Martins, M.~A. Dahleh, and N.~Elia, ``{Feedback stabilization of
  uncertain systems in the presence of a direct link},'' \emph{\IEEEJAC},
  51: 438--447, 2006.

\bibitem{Hayakawa2009}
T.~Hayakawa, H.~Ishii, and K.~Tsumura, ``{Adaptive quantized control for linear
  uncertain discrete-time systems},'' \emph{Automatica}, 45: 692--700, 2009.

\bibitem{Vu2012}
L.~Vu and D.~Liberzon, ``{Supervisory control of uncertain systems with
  quantized information},'' \emph{\IJACSP}, 26: 739--756, 2012.

\bibitem{Elia2001}
N.~Elia and S.~K.~Mitter, 
``Stabilization of linear systems with limited information,'' 
\emph{\IEEEJAC}, 46: 1384--1400, 2001.

\bibitem{Okano2012b}
K.~Okano and H.~Ishii, ``{Data rate limitations for stabilization of uncertain
  systems},'' in \emph{Proc. of the 51st IEEE Conference on Decision and
  Control}, 2012, pp.~3286--3291.

\bibitem{Okano2014a}
------, ``{Stabilization of uncertain systems with finite data rates and
  Markovian packet losses},'' \emph{IEEE Trans. Control of Network Syst.},
  1: 298-307, 2014.

\bibitem{Kang2015}%
X.~Kang and H.~Ishii, 
``Coarsest quantization for networked control of uncertain linear systems,'' 
\emph{Automatica}, 51: 1--8, 2015. 

\bibitem{Okano2014b}
K.~Okano and H.~Ishii, ``{Stabilization of uncertain systems using quantized
  and lossy observations and uncertain control inputs},'' in \emph{Proc.
  of European Control Conference}, 2014, pp. 240--245.

\bibitem{Liberzon2005}
D.~Liberzon and J.~P.~Hespanha, ``Stabilization of nonlinear systems with
 limited information feedback,'' \emph{\IEEEJAC}, 50: 910--915, 2005.

\bibitem{Li2004}
K.~Li and J.~Baillieul, ``{Robust quantization for digital finite communication
  bandwidth (DFCB) control},'' \emph{\IEEEJAC}, 49: 1573--1584, 2004.

\bibitem{Tsumura2009a}
K.~Tsumura, ``{Optimal quantization of signals for system identification},''
  \emph{\IEEEJAC}, 54: 2909--2915, 2009.

\bibitem{Moore2009}
R.~E. Moore, R.~B. Kearfott, and M.~J. Cloud, \emph{{Introduction to Interval
  Analysis}}.\hskip 1em plus 0.5em minus 0.4em\relax Philadelphia: SIAM, 2009.

\bibitem{Xie2009}
L.~Xie and L.~Xie, ``{Stability analysis of networked sampled-data linear
  systems with Markovian packet losses},'' \emph{\IEEEJAC}, 54: 1375--1381, 2009.

\bibitem{Costa2005}
O.~L.~V. Costa, M.~D. Fragoso, and R.~P. Marques, \emph{{Discrete-Time Markov
  Jump Linear Systems}}. %
  \hskip 1em plus 0.5em minus 0.4em\relax 
  London: Springer, 2005.
\end{thebibliography}
\end{document}